\documentclass[preprint,floatfix,aps,nofootinbib]{revtex4}
\usepackage{graphicx}
\usepackage{dcolumn}
\usepackage{bm}
\usepackage{amsmath,amssymb}

\newcommand {\be}  {\begin{equation}}   
\newcommand {\ee}  {\end{equation}}
\newcommand {\bee} {\begin{equation*}}   
\newcommand {\eee} {\end{equation*}}
\newcommand {\bea} {\begin{eqnarray}}   
\newcommand {\eea} {\end{eqnarray}}
\newcommand {\beaa}{\begin{eqnarray*}}   
\newcommand {\eeaa}{\end{eqnarray*}}
\newcommand {\bse} {\begin{subequations}}
\newcommand {\ese} {\end{subequations}}
\newcommand {\nn}  {\nonumber}
\newcommand {\pr}  {\prime}
\newcommand {\pd}  {\partial}
\newcommand {\sig} {\sigma}
\newcommand {\onehalf}{\frac{1}{2}}
\newcommand {\vect}[1]{\mathbf{#1}}
\newcommand {\bx}   {\vect{x}}
\newcommand {\bD}   {\vect{D}}
\newcommand {\bV}   {\vect{V}}
\newcommand {\bID}   {\vect{I_D}}
\newcommand {\bIV}   {\vect{I_V}}

\begin{document}
\title{Analytic physical model of anisotropic anomalous diffusion}

\author{Billy D. Jones} \email{bjones@apl.uw.edu}
\affiliation{Applied Physics Laboratory, University of Washington, Seattle, WA 98105}
\date{Draft of 2013-06-26}

\begin{abstract} 

The temporal Fokker-Planck equation is analytically integrated in an arbitrary number of spatial dimensions but 
with the 2D and 3D cases highlighted. It is shown that a temporal power-law ansatz for the anisotropic diffusion coefficients 
leads naturally to a physically reasonable model of normal and anomalous diffusion with drift. 
An interpretation of the drift and diffusion coefficients is provided and the analytic 
growth rate of the area and volume of uncertainty is determined. It is shown that 
the asymptotic growth of the uncertainty in the volume of space about the mean position
is proportional to the square-root of time raised to the 
power of the sum over all exponents of the anisotropic diffusion coefficients.

\end{abstract}
\maketitle

\section{Introduction}

Normal and anomalous diffusion models have been used in such diverse topics as 
submarine search \cite{Krout1, Shlesinger}; Brownian motion; 
disease, animal, and cell movement \cite{rwbio}; 
ballistic or wavelike movement; turbulent diffusion; 
and particle acceleration \cite{supradiff}. The current paper provides an 
analytic solution to the temporal Fokker-Planck equation and shows how a physical model of anomalous diffusion  
naturally arises given an ansatz of temporal power-law growth in the anisotropic diffusion coefficients.

\section{Analytic temporal Fokker-Planck}
\label{sec:analytic} 

For clarity, we start in one spatial dimension, but the results generalize 
to an arbitrary number of dimensions as will be shown. 
The Fokker-Planck (FP) equation in 1D with spatially constant but temporally varying diffusion and drift coefficients (i.e.\ the 
`temporal' FP equation) is given by 
\be
\frac{\pd \rho(x,t)}{\pd t}=\frac{D(t)}{2}\frac{\pd^2\rho(x,t)}{\pd x^2}-V(t)\frac{\pd\rho(x,t)}{\pd x}\;,
\label{eq:fp1d}
\ee
where $\rho(x,t)$ is a normalized probability density of interest (e.g.\ for a submarine, dust, biologic, or financial particle):
\be
\int_{-\infty}^\infty\rho(x,t)dx=1\;,
\ee
$D(t)>0$ is the diffusion coefficient, and $V(t)\in\mathbb{R}$ is the drift coefficient. 
Both $D(t)$ and $V(t)$ have further temporal integrability conditions as discussed below.
Simple dimensional analysis applied to Eq.~(\ref{eq:fp1d}) gives 
\bse
\label{eq:diman}
\bea
[D(t)] &=& \left[\frac{\pd x^2}{\pd t}\right]\Longrightarrow {\rm area/time}\;\\
&{\rm and}&\nn\\
{[V(t)]} &=& \left[\frac{\pd x}{\pd t}\right]\Longrightarrow {\rm distance/time}\;.
\eea
\ese
The latter is a drift velocity and the former appears to be an area growth rate. 
This intuition is shown to be correct in 2D and the N-dimensional generalization is developed.

With initial conditions at time $t=t_0$ set 
to be a Gaussian with mean position $x_0$ and variance about the mean $\sig^2_{0x}$, 
a solution (verified analytically below) to Eq.~(\ref{eq:fp1d}) is given by
\be
\rho(x,t)={\cal N}\left[x;\,x_0+I_{V}(t),\,\sig^2_{0x}+I_{D}(t)\right]\;,
\label{eq:soln1d}
\ee
where $I_{V}(t)$ and $I_{D}(t)$ are integrals of the drift and diffusion coefficients respectively:
\bse
\label{eq:IvId}
\bea
I_{V}(t)&\equiv&\int_{t_0}^tV(t')dt'\\
&{\rm and}&\nn\\
I_{D}(t)&\equiv&\int_{t_0}^tD(t^\pr)dt'\;,
\eea
\ese
and ${\cal N}[x;a(t),b(t)]$ is standard notation for a normalized Gaussian with mean $a(t)$ and 
variance $b(t)$: 
\be
{\cal N}\left[x;\,a(t),b(t)\right]=\frac{e^{-\frac{\left[x-a(t)\right]^2}{2\,b(t)}}}{\sqrt{2\pi b(t)}}\;.
\label{eq:abN}
\ee
The position argument $x$ is added for clarity and to allow for easier generalization to higher dimensions; 
the parameters $a(t)$ and $b(t)$ depend on time but not space. 
The integrals of Eq.~(\ref{eq:IvId}) of course have to converge for this solution to make sense 
and their derivatives give back the respective integrand exactly:
\bse
\label{eq:DIvId}
\bea
\frac{\pd I_{V}(t)}{\pd t}&=&V(t)\\
&{\rm and}&\nn\\
\frac{\pd I_{D}(t)}{\pd t}&=&D(t)\;;
\eea
\ese
these relations especially will be used in the analytic proof that follows. 
Also note that by definition these integral definitions vanish at the initial condition point:
\bse
\label{eq:icIvId}
\bea
I_{V}(t_0)&=&0\;,\\
I_{D}(t_0)&=&0\;.
\eea
\ese
Finally, notice how the units are correct and a time integral of $V(t)$ and $D(t)$ produce 
quantities of dimension length and area respectively, consistent with Eqs.~(\ref{eq:diman}) and (\ref{eq:soln1d}).

Calculating the mean position and variance about the mean at arbitrary time $t$ gives
\bea
\left<x\right> &\equiv& \int_{-\infty}^\infty\rho(x,t)\,x\,dx=\boxed{x_0+I_{V}(t)}\\
&{\rm and}&\nn\\ 
(\Delta x)^2&\equiv&\int_{-\infty}^\infty\rho(x,t)\left(x-\left<x\right>\right)^2dx=\boxed{\sig^2_{0x}+I_{D}(t)}
\eea
respectively. At the initial time, $t=t_0$, these collapse to the Gaussian initial conditions, 
and for all time we see that these integrals of Eq.~(\ref{eq:IvId}) are the required additions 
to the exact solution of a drifting and diffusing Gaussian.  
We see that a Gaussian remains a Gaussian for all time here, 
but the solution is not stationary: both the mean and variance depend on time. This time 
dependence is modeled as a power law below and this is where the anomalous diffusion lies. 

\subsection{Analytic proof}
\label{sec:proof}

Now we analytically show that $\rho(x,t)$ of Eq.~(\ref{eq:soln1d}) is an exact solution to Eq.~(\ref{eq:fp1d}). 
Writing out the solution more explicitly from above 
gives 
\bse
\label{eq:abeqn}
\bea
\rho(x,t)&=&\frac{e^{-\frac{\left[x-a(t)\right]^2}{2\,b(t)}}}{\sqrt{2\pi b(t)}}\\
&{\rm where}&\nn\\ 
a(t)&=&x_0+I_{V}(t)\;,\\
b(t)&=&\sig^2_{0x}+I_{D}(t)\;.
\eea
\ese
For the proof that follows, note that the exact temporal derivatives of $a(t)$ and $b(t)$ are given by
\bse
\label{eq:ab}
\bea
\frac{\pd a(t)}{\pd t}&=&\frac{\pd I_{V}(t)}{\pd t}=V(t)\\
&{\rm and}&\nn\\ 
\frac{\pd b(t)}{\pd t}&=&\frac{\pd I_{D}(t)}{\pd t}=D(t)
\eea
\ese
respectively, where $V(t)$ and $D(t)$ are the same coefficients as in Eq.~(\ref{eq:fp1d}). 
We have not specified the temporal dependence of $V(t)$ or $D(t)$ yet, other than requiring that they are both integrable 
from some initial time $t_0$ through an arbitrarily large but finite final time $t$. 
However, in the next section their temporal dependence is discussed at length because that is 
what determines if the diffusion is anomalous or not. 

Taking a partial time derivative of Eq.~(\ref{eq:abeqn}) using the product rule of differentiation and the 
results of Eq.~(\ref{eq:ab}) gives
\be
\label{eq:A}
\frac{\pd \rho(x,t)}{\pd t}=\left\{\frac{V(t)\left[x-a(t)\right]}{b(t)}+\frac{D(t)
\left[x-a(t)\right]^2}{2\,b^2(t)}-\frac{D(t)}{2\,b(t)}\right\}\rho(x,t)\;.
\ee
Likewise, the first and second partial spatial derivatives of $\rho(x,t)$ are 
\be
\frac{\pd \rho(x,t)}{\pd x}=-\frac{\left[x-a(t)\right]}{b(t)}\rho(x,t)
\ee
and
\be
\frac{\pd^2 \rho(x,t)}{\pd x^2}=\left\{\frac{\left[x-a(t)\right]^2}{b^2(t)}-\frac{1}{b(t)}\right\}\rho(x,t)
\ee
respectively; multiplying the latter by $D(t)/2$ and the former  
by $-V(t)$ according to the right-hand side of Eq.~(\ref{eq:fp1d}) gives 
\be
\label{eq:AA}
\frac{D(t)}{2}\frac{\pd^2\rho(x,t)}{\pd x^2}-V(t)\frac{\pd\rho(x,t)}{\pd x}=
\left\{\frac{D(t)\left[x-a(t)\right]^2}{2\,b^2(t)}-\frac{D(t)}{2\,b(t)}+
\frac{V(t)\left[x-a(t)\right]}{b(t)}\right\}\rho(x,t)\;.
\ee
Comparing this with the time-partial result, Eq.~(\ref{eq:A}), 
shows that they are equivalent and therefore Eq.~(\ref{eq:fp1d}) is 
satisfied exactly for $\rho(x,t)$ of Eq.~(\ref{eq:soln1d})---q.e.d. 

\subsection{Multiple spatial dimensions}
\label{sec:multidim}

Now we generalize the analytic solution of the previous section to multiple spatial dimensions. 
Due to the linearity of Eq.~(\ref{eq:fp1d}), this is easily accomplished with a product of Gaussians as will be shown. 
The solution easily generalizes to an arbitrary number of spatial dimensions, however the 2D and 3D solutions are 
emphasized next due to their physical relevance. The N-dimensional results are presented near the end of  
the paper. We generalize Eq.~(\ref{eq:fp1d}) with the 
anisotropic application in mind, hence we introduce diffusion coefficients 
$D_x(t)$, $D_y(t)$, and $D_z(t)$. These diffusion coefficients are in general of different magnitudes, 
especially with the vertical $D_z(t)$ often smaller or much smaller than its 
horizontal counterparts often themselves of comparable magnitude: $D_x(t)\sim D_y(t)$. 
For completeness note that the drift coefficients, $\bV(t)$, are also treated anisotropically in what follows.

\subsubsection{Analytic anisotropic 2D temporal Fokker-Planck}
\label{sec:2d}

Next, we generalize the above temporal FP equation to two spatial dimensions  
and discuss the interpretation of its anisotropic diffusion and drift coefficients.  
Eq.~(\ref{eq:fp1d}) generalized to 2D with the diffusion and drift coefficients allowed 
to be anisotropic is    
\be
\label{eq:fp2d}
\frac{\pd \rho(x,y,t)}{\pd t}=\onehalf\left[D_x(t)\frac{\pd^2}{\pd x^2}+D_y(t)\frac{\pd^2}{\pd y^2}\right]\rho(x,y,t)
-\left[V_x(t)\frac{\pd}{\pd x}+V_y(t)\frac{\pd}{\pd y}\right]\rho(x,y,t)
\;.
\ee
Maintaining the notation defined in the previous section, a solution to this 2D equation is simply a product of normalized Gaussians: 
\bse
\label{eq:soln2d}
\bea
\rho(x,y,t)&=&
{\cal N}\left[x;\,x_0+I_{V_x}(t),\,\sig^2_{0x}+I_{D_x}(t)\right]\,
{\cal N}\left[y;\,y_0+I_{V_y}(t),\,\sig^2_{0y}+I_{D_y}(t)\right]\\
&\equiv&{\cal N}_x\,{\cal N}_y\;.
\eea
\ese
Especially note this definition of ${\cal N}_x$ and ${\cal N}_y$ in this last line; this is nothing more than a shorthand used below, but 
the full expression of the previous line is implied. 
In this solution, the anisotropic 2D drift and diffusion coefficient time integrals are defined by 
\bse
\label{eq:IvId2d}
\bea
I_{V_x}(t)&\equiv&\int_{t_0}^tV_x(t')dt'\;,\\
I_{V_y}(t)&\equiv&\int_{t_0}^tV_y(t')dt'\;,\\
I_{D_x}(t)&\equiv&\int_{t_0}^tD_x(t^\pr)dt'\;,\\
&{\rm and}&\nn\\
I_{D_y}(t)&\equiv&\int_{t_0}^tD_y(t^\pr)dt'\;.
\eea
\ese
Due to the product rule of differentiation and the linearity of the derivatives in Eq.~(\ref{eq:fp2d}), 
this product of Gaussians is easily seen to be a solution. Explicitly we have for the temporal derivative:
\be
\frac{\pd \rho(x,y,t)}{\pd t}=
\frac{\pd}{\pd t}\left({\cal N}_x\,{\cal N}_y\right)=\left(\frac{\pd}{\pd t}{\cal N}_x\right){\cal N}_y+{\cal N}_x\left(\frac{\pd}{\pd t}{\cal N}_y\right)\;,
\ee
and for the spatial derivatives:
\bse
\bea
\frac{\pd \rho(x,y,t)}{\pd x}&=&
\frac{\pd}{\pd x}\left({\cal N}_x\,{\cal N}_y\right)=\left(\frac{\pd}{\pd x}{\cal N}_x\right){\cal N}_y\;,\\
\frac{\pd \rho(x,y,t)}{\pd y}&=&
\frac{\pd}{\pd y}\left({\cal N}_x\,{\cal N}_y\right)={\cal N}_x\left(\frac{\pd}{\pd y}{\cal N}_y\right)\;,\\
\frac{\pd^2 \rho(x,y,t)}{\pd x^2}&=&
\frac{\pd^2}{\pd x^2}\left({\cal N}_x\,{\cal N}_y\right)=\left(\frac{\pd^2}{\pd x^2}{\cal N}_x\right){\cal N}_y\;,\\
\frac{\pd^2 \rho(x,y,t)}{\pd y^2}&=&
\frac{\pd^2}{\pd y^2}\left({\cal N}_x\,{\cal N}_y\right)={\cal N}_x\left(\frac{\pd^2}{\pd y^2}{\cal N}_y\right)\;.
\eea
\ese
Thus, combining all the pieces of Eq.~(\ref{eq:fp2d}), this product of Gaussians solution satisfies 
\be
{\cal N}_y\left(\frac{\pd}{\pd t}-\frac{D_x(t)}{2}\frac{\pd^2}{\pd x^2}+V_x(t)\frac{\pd}{\pd x}\right){\cal N}_x+
{\cal N}_x\left(\frac{\pd}{\pd t}-\frac{D_y(t)}{2}\frac{\pd^2}{\pd y^2}+V_y(t)\frac{\pd}{\pd y}\right){\cal N}_y=0
\;,
\ee
and we see that the equation factorizes into two {\it independent} 1D equations. Thus we can 
use the 1D results of the previous section and Eq.~(\ref{eq:soln2d}) is seen to be a solution of Eq.~(\ref{eq:fp2d})---q.e.d.

Calculating the 2D mean position and variance about the mean at arbitrary time $t$, 
we see that they factorize along with the solution and its norm. Explicitly, we have  
\bse
\bea
1&=&\int_{-\infty}^\infty dx\int_{-\infty}^\infty dy\,\rho(x,y,t)=\int_{-\infty}^\infty {\cal N}_x\,dx\int_{-\infty}^\infty {\cal N}_y\,dy\;,\\
\left<x\right> &\equiv& \int_{-\infty}^\infty dx\int_{-\infty}^\infty dy\,\rho(x,y,t)\,x=x_0+I_{V_x}(t)\;,\\
\left<y\right> &\equiv& \int_{-\infty}^\infty dx\int_{-\infty}^\infty dy\,\rho(x,y,t)\,y=y_0+I_{V_y}(t)\;,\\
(\Delta x)^2&\equiv&\int_{-\infty}^\infty dx\int_{-\infty}^\infty dy\,\rho(x,y,t)\left(x-\left<x\right>\right)^2=\sig^2_{0x}+I_{D_x}(t)\;,\\
(\Delta y)^2&\equiv&\int_{-\infty}^\infty dx\int_{-\infty}^\infty dy\,\rho(x,y,t)\left(y-\left<y\right>\right)^2=\sig^2_{0y}+I_{D_y}(t)\;.
\eea
\ese
Writing these results in vector notation (here 2-vectors---in the next section, same notation but they will be 3-vectors) gives
\bea
\label{eq:meanX}
\left<\bx\right> &\equiv& \int \rho(\bx,t)\,\bx\,d^2x=\boxed{\vect{x}_0+\bIV(t)}\\
&{\rm and}&\nn\\ 
\label{eq:deltaX2}
(\Delta \bx)^2&\equiv&\int\rho(\bx,t)\left(\bx-\left<\bx\right>\right)^2\,d^2x=
\boxed{\mbox{\boldmath$\sig$}_0^2+\bID(t)\cdot\left(\hat{x}+\hat{y}\right)}\;,
\eea
for Cartesian unit vectors $\hat{x}$ and $\hat{y}$, 
where the drift and diffusion coefficient time integrals are here 2-vectors: 
\bse
\bea
\bIV(t)&\equiv&\left[I_{V_x}(t),I_{V_y}(t)\right]\;,\\
\bID(t)&\equiv&\left[I_{D_x}(t),I_{D_y}(t)\right]\;,\\
&{\rm and}&\nn\\ 
\bID(t)\cdot\left(\hat{x}+\hat{y}\right)&\equiv&I_{D_x}(t)+I_{D_y}(t)\;.
\eea
\ese

Continuing with the interpretation of the 2D drift and diffusion coefficients. 
First the drift: 
The drift coefficient vector, $\bV(t)$, is equivalent to the 
instantaneous velocity of the mean position. This follows from taking a derivative of 
Eq.~(\ref{eq:meanX}): 
\be
\frac{\pd\left<\bx\right>}{\pd t}=\frac{\pd\bIV(t)}{\pd t}=\bV(t)=\left[V_x(t),V_y(t)\right]\;.
\label{eq:meanXrate}
\ee
In summary: the drift coefficient vector $\bV(t)$ is equal to the instantaneous velocity of the mean position 
of the probability density of interest. This interpretation is still valid in an arbitrary number of spatial dimensions 
with arbitrary time-integrable anisotropic drift coefficients. 

Second the diffusion: 
For the interpretation of the diffusion coefficient vector, first calculate the area of uncertainty (AOU) of 
the 2D variance about the mean position at arbitrary time $t$. Using the $1\sig$ ellipse for its definition, we have 
\be
AOU(t)=\pi\Delta x\Delta y=\pi\sqrt{\sig^2_{0x}+I_{D_x}(t)}\sqrt{\sig^2_{0y}+I_{D_y}(t)}\;.
\ee
Taking a time derivative for the AOU growth rate gives
\bea
\frac{\pd AOU(t)}{\pd t}&=&\pi\left[\frac{D_x(t)}{2\sqrt{\sig^2_{0x}+I_{D_x}(t)}}\sqrt{\sig^2_{0y}+I_{D_y}(t)}+
\sqrt{\sig^2_{0x}+I_{D_x}(t)}\frac{D_y(t)}{2\sqrt{\sig^2_{0y}+I_{D_y}(t)}}\right]\nn\\
&=&\boxed{\frac{AOU(t)}{2}\left[\frac{D_x(t)}{\sig^2_{0x}+I_{D_x}(t)}+
\frac{D_y(t)}{\sig^2_{0y}+I_{D_y}(t)}\right]}
\label{eq:aourate}
\;,
\eea
a major result, so we box it. The interpretation of the diffusion coefficient is particularly simple 
in the 2D isotropic case for then we have ($D_x(t)=D_y(t)\equiv D(t)$, $\sig^2_{0x}=\sig^2_{0y}\equiv\sig^2_{0}$): 
\be
\frac{\pd AOU(t)}{\pd t}{\Biggr{|}}_{\rm isotropic\;2D}=AOU(t)\frac{D(t)}{\sig^2_{0}+I_{D}(t)}=\boxed{\pi D(t)}\;. 
\label{eq:niceone}
\ee
Thus, except for the inconsequential factor of $\pi$, the 2D isotropic diffusion coefficient is equal to the 
instantaneous AOU growth rate---certainly the units are right as we discussed earlier, and here we showed 
this intuition to be precisely correct in 2D. Below we see this precise interpretation breaks down 
in higher spatial dimensions---turns out the 2D case is somewhat of an accident, as will be shown. 

\subsubsection{Analytic anisotropic 3D temporal Fokker-Planck}
\label{sec:3d}

It is clear that the 3D case (and in general N-dimensional case) of the solution under discussion generalizes 
just like the 2D case of the previous section. 
But for completeness, we will write out some of the 3D results explicitly here. First, the anisotropic 3D FP equation with 
spatially constant but temporally varying diffusion and drift coefficients is 
\be
\frac{\pd \rho(\bx,t)}{\pd t}=\onehalf\bD(t)\cdot\nabla^2\rho(\bx,t)-\bV(t)\cdot\nabla\rho(\bx,t)\;,
\label{eq:fp3d}
\ee
where the generalized Laplacian is defined by 
\be
\bD(t)\cdot\nabla^2\equiv D_x(t)\frac{\pd^2}{\pd x^2}+D_y(t)\frac{\pd^2}{\pd y^2}+D_z(t)\frac{\pd^2}{\pd z^2}
\ee
and the drift advection operator is given by 
\be
\bV(t)\cdot\nabla=V_x(t)\frac{\pd}{\pd x}+V_y(t)\frac{\pd}{\pd y}+V_z(t)\frac{\pd}{\pd z}\;. 
\ee
Using the notation of the previous section (with 3-vectors instead of 2-vectors), 
and in particular using the shorthand of Eq.~(\ref{eq:soln2d}) (with the addition of a shorthand for ${\cal N}_z$), the 
solution, norm, mean, and variance in 3D is given by 
\bea
\rho(\bx,t)&=&{\cal N}_x\,{\cal N}_y\,{\cal N}_z\;,\\
1&=&\int\rho(\bx,t)d^3x\;,\\
\left<\bx\right> &=& \int \rho(\bx,t)\,\bx\,d^3x=\boxed{\vect{x}_0+\bIV(t)}\;,\label{eq:39}\\
&{\rm and}&\nn\\
(\Delta \bx)^2&=&\int\rho(\bx,t)\left(\bx-\left<\bx\right>\right)^2\,d^3x=\boxed{\mbox{\boldmath$\sig$}_0^2+
\bID(t)\cdot\left(\hat{x}+\hat{y}+\hat{z}\right)}
\label{eq:40}
\eea
respectively. 
The proof that this is a solution of the above 3D temporal FP equation follows almost identically to the 2D case of the previous section 
and will not be repeated here. 

We close this section by discussing the interpretation of the drift and diffusion coefficients in 3D. 
First, the drift coefficient vector, $\bV(t)$, follows identically to Eq.~(\ref{eq:meanXrate}) of the 2D case except that now 
it is a 3-vector: 
\be
\frac{\pd\left<\bx\right>}{\pd t}=\frac{\pd\bIV(t)}{\pd t}=\bV(t)=\left[V_x(t),V_y(t),V_z(t)\right]
\;.\label{eq:vxyz}
\ee
Second, the diffusion coefficient vector, $\bD(t)$, follows similarly to the above except now it is a volume instead of area 
of uncertainty. (Note for later: In $N>3$ spatial dimensions we keep calling it `volume'.) 
Using the $1\sig$ ellipsoid for the definition of the volume of uncertainty (VOU) in three dimensions, we have 
\be
VOU(t)=\frac{4}{3}\pi\Delta x\Delta y\Delta z=\frac{4}{3}\pi\sqrt{\sig^2_{0x}+I_{D_x}(t)}\sqrt{\sig^2_{0y}+I_{D_y}(t)}\sqrt{\sig^2_{0z}+I_{D_z}(t)}\;.
\ee
Similar to Eq.~(\ref{eq:aourate}), but for volume instead of area, taking a time derivative gives the VOU growth rate (skipping similar 
algebra to the 2D case): 
\be
\frac{\pd VOU(t)}{\pd t}
=\boxed{\frac{VOU(t)}{2}\left[\frac{D_x(t)}{\sig^2_{0x}+I_{D_x}(t)}+
\frac{D_y(t)}{\sig^2_{0y}+I_{D_y}(t)}+\frac{D_z(t)}{\sig^2_{0z}+I_{D_z}(t)}\right]}
\label{eq:vourate}
\;,
\ee
the same form as in 2D even with the same factor of $2$ in the 
denominator---this factor of $2$ leads to an accident in 2D---but now there are three terms 
(and in N dimensions, N terms). As shown below, these N terms lead to an overall multiplicative factor of N in the 
exponent of the asymptotic temporal growth of the volume uncertainty in the isotropic case. 
This will become clear below after we discuss the temporal dependence of 
the diffusion and drift coefficients and show how they relate to the different types  
of normal and anomalous diffusion. 

\section{Anisotropic anomalous diffusion}
\label{sec:anomalous} 

The analytic solution of the previous section is developed with a power-law ansatz for  
the diffusion coefficient. Anomalous diffusion is seen to arise and the 
physical bounds of the diffusion coefficient exponents are discussed. In the final section, the volume growth rate 
is analyzed in the general case of N spatial dimensions. This generalizes the above 2D result 
of the AOU growth rate being identified with the diffusion coefficient (modulo a factor of $\pi$). 

\subsection{Power-law ansatz}

The drift coefficient vector, $\bV(t)$, is the instantaneous velocity of the mean position as discussed 
above around  Eqs.~(\ref{eq:meanXrate}) and (\ref{eq:vxyz}) in 2D and 3D respectively. 
The diffusion coefficient vector, $\bD(t)$, affects the centered variance, $(\Delta \bx)^2$, 
according to Eq.~(\ref{eq:40}), but note that the drift $\bV(t)$ decouples from this diffusion.
It is true that the mean squared position does depend on $\bV(t)$ as follows from 
Eqs.~(\ref{eq:39}) and (\ref{eq:40}): 
\be
\left<\bx^2\right>=(\Delta \bx)^2+\left<\bx\right>^2=
\mbox{\boldmath$\sig$}_0^2+\bID(t)\cdot\left(\hat{x}+\hat{y}+\hat{z}\right)+\left[\vect{x}_0+\bIV(t)\right]^2
\;.
\ee
However, the {\it centered} variance $(\Delta \bx)^2$ is what matters with regards to 
whether the diffusion is anomalous or not as we now show.

To study anomalous diffusion, let the components of $\bD(t)$ be given by a temporal power law with 
power $\alpha=1$ matched to the normal-diffusion case. Thus, in general for anisotropic diffusion, let 
\be
\bD(t)\equiv\left[C_x\,\alpha_x\,(t-t_0)^{\alpha_x-1},C_y\,\alpha_y\,(t-t_0)^{\alpha_y-1},C_z\,\alpha_z\,(t-t_0)^{\alpha_z-1}\right]
\;,\label{eq:powerlaw}
\ee
for constant vectors $\vect{C}$ and \mbox{\boldmath$\alpha$}. 
Plugging this power-law ansatz into Eq.~(\ref{eq:40}) for the centered variance, and 
performing the time integrations gives
\be
(\Delta \bx)^2=\boxed{\mbox{\boldmath$\sig$}_0^2+C_x\,(t-t_0)^{\alpha_x}+C_y\,(t-t_0)^{\alpha_y}+C_z\,(t-t_0)^{\alpha_z}}
\label{eq:powerlawint}
\;,
\ee
{\it as long as} the exponents are all positive:
 $\alpha_x>0$, $\alpha_y>0$, {\it and} $\alpha_z>0$; for zero exponents there is a logarithmic divergence 
at initial time $t=t_0$ and for negative exponents there is a power-law divergence. 
Thus we see that subdiffusion ($\alpha<1$) is allowed in this model, 
but the exponents must be positive ($\alpha>0$) in order for the centered variance to be finite as 
required on physical grounds \cite{anomdiff}. 

For superdiffusion ($\alpha>1$), the ballistic velocity,  
\be
{\rm ballistic\,velocity}\equiv\frac{(\Delta \bx)^2}{\left(t-t_0\right)^2}\stackrel{t\gg t_0}{\sim}
C_x\,t^{\alpha_x-2}+C_y\,t^{\alpha_y-2}+C_z\,t^{\alpha_z-2}\;,
\ee
in the asymptotic large-time limit seems to require 
$\alpha_x\leq 2$, $\alpha_y\leq 2$, {\it and} $\alpha_z\leq 2$ for finiteness. However, 
there is no reason in general to limit this upper bound of the diffusion coefficient exponents: 
The principle of relativity already naturally bounds the exponents as nicely explained in \cite{supradiff}.  
Accelerating particles for short times have $\alpha=3$ (turbulent diffusion) or even $\alpha=4$ (nonrelativistic acceleration), 
however for long times relativity bounds these exponents to the ballistic $\alpha=2$ limit (see \cite{supradiff}).

\begin{figure}[ht]
\includegraphics[width=\textwidth]{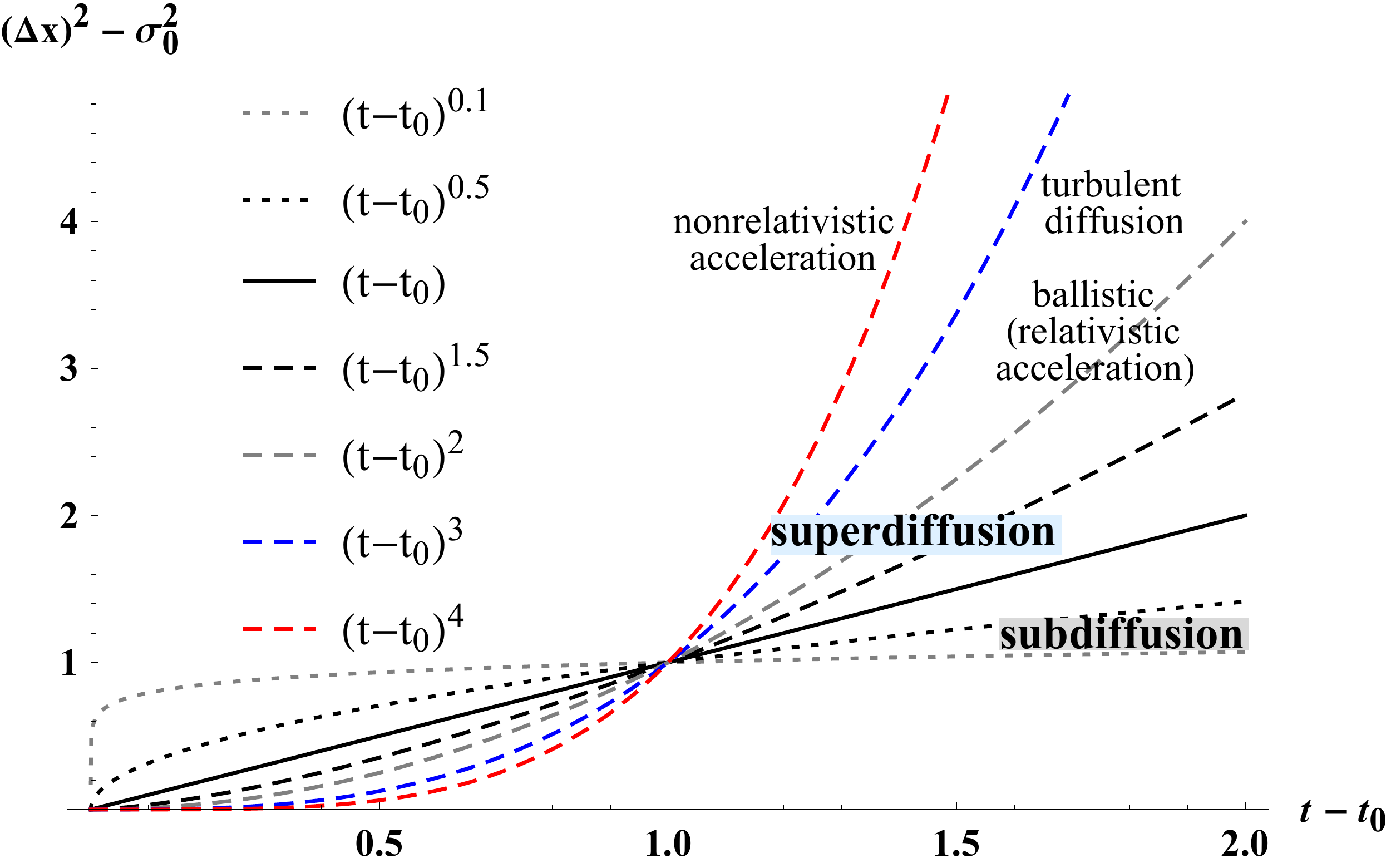}
\caption{Variance of the position about the mean from integrating the power-law ansatz of the current paper 
[see Eqs.~(\ref{eq:powerlaw}) and (\ref{eq:powerlawint})]. The isotropic case is shown for 
simplicity (in arbitrary units): $(\Delta \bx)^2-\mbox{\boldmath$\sig$}_0^2\sim\left(t-t_0\right)^\alpha$. 
$\mbox{\boldmath$\sig$}_0^2$ is the variance of the initial Gaussian at time $t=t_0$. 
Normal diffusion has $\alpha=1$; subdiffusion has 
$0<\alpha<1$; and superdiffusion has $\alpha>1$ \cite{anomdiff}. The turbulent diffusion and particle acceleration 
solutions are explained in \cite{supradiff}.}
\label{fig:anomdiff}
\end{figure}

In summary, the power-law ansatz of Eq.~(\ref{eq:powerlaw}) implies Eq.~(\ref{eq:powerlawint}) for the 
centered position variance. For simplicity, in the isotropic case this is 
$(\Delta \bx)^2-\mbox{\boldmath$\sig$}_0^2\sim\left(t-t_0\right)^\alpha$, 
where normal diffusion has $\alpha=1$, subdiffusion has $0<\alpha<1$, and superdiffusion has $\alpha>1$ \cite{anomdiff}. 
Figure~\ref{fig:anomdiff} 
shows these different phases of anomalous diffusion 
as follows from integrating the power-law ansatz of 
the analytic model of the current paper.

\subsection{Volume growth rate in N spatial dimensions}

The power law of the previous section is used to discuss the asymptotic large-time growth rate of the 
N-dimensional `volume' of uncertainty (VOU).  
Given semi-axis radii of the respective $1\sigma$ variance `ellipsoid', the VOU in N spatial dimensions 
is given by $VOU=2a$ for $N=1$; $VOU=\pi a\,b$ for $N=2$; $VOU=\frac{4}{3}\pi a\,b\,c$ 
for $N=3$; $VOU=\frac{\pi^2}{2} a\,b\,c\,d$ for $N=4$; etc.
Eq.~(\ref{eq:vourate}) generalized to N spatial dimensions is  
\be
\frac{\pd VOU(t)}{\pd t}
=\boxed{\frac{VOU(t)}{2}\sum_{i=1}^{i=N}\frac{D_{x_i}(t)}{\sig^2_{0x_i}+I_{D_{x_i}}(t)}}
\label{eq:vourateNdim}
\;.
\ee
Plugging in the power-law ansatz of Eqs.~(\ref{eq:powerlaw}) and (\ref{eq:powerlawint}), 
in the large-time asymptotic limit, we see these constant vectors $\vect{C}$ cancel in the numerator 
and denominator and we are left with 
\bea
\frac{\pd VOU(t)}{\pd t}
&\stackrel{t\gg t_0}{\sim}&\frac{VOU(t)}{2}\sum_{i=1}^{i=N}\frac{C_{x_i}\,\alpha_{x_i}\,(t-t_0)^{\alpha_{x_i}-1}}{C_{x_i}\,(t-t_0)^{\alpha_{x_i}}}\nn\\
&\stackrel{t\gg t_0}{\sim}&\frac{VOU(t)}{2\,t}\sum_{i=1}^{i=N}\alpha_{x_i}\nn\\
&\equiv&\boxed{\frac{VOU(t)}{2\,t}\alpha_{N\!sum}}
\label{eq:vourateNdimAsym}
\;,
\eea
where $\alpha_{N\!sum}$ is a shorthand for the sum of the power-law exponents over all N spatial dimensions. This asymptotic equation 
is easy to integrate and we are left with
\be
\boxed{VOU(t)\stackrel{t\gg t_0}{\sim}t^{\frac{\mbox{\small$\alpha$}_{\!N\!sum}}{2}}}\;.
\label{eq:niceoneN}
\ee
This $2$ in the denominator of the exponent in the last formula leads to an accident for spatial dimension $N=2$, 
leading to a nice interpretation for the 2D isotropic diffusion coefficient: Eq.~(\ref{eq:niceone}); 
however, this interpretation is not valid for $N\neq2$, and Eq~(\ref{eq:vourateNdimAsym}) 
or its integral Eq.~(\ref{eq:niceoneN}) must in general be used. Eq.~(\ref{eq:niceoneN}) described in words: 
The uncertainty in the volume of N-dimensional space about the mean position  
grows like the square-root of time raised to the power of the sum over all 
exponents of the anisotropic diffusion coefficients. This is the same square-root of time as 
Einstein's Brownian motion, and the $\alpha_{N\!sum}$ generalizes the result to N spatial dimensions 
with anisotropic anomalous diffusion.

\section{Summary and Discussion}

The temporal Fokker-Planck equation was defined and analytically integrated in an arbitrary number of dimensions, but 
with the 2D and 3D cases highlighted. It was shown that a power-law ansatz for the anisotropic diffusion coefficients leads 
to a natural physical model of anomalous diffusion with drift. The volume of uncertainty was determined for an arbitrary 
number of spatial dimensions N to grow like (e.g.\ in the isotropic case): $VOU(t)\sim t^{N\alpha/2}$, where $\alpha$ is the 
temporal exponent of the power-law ansatz for the diffusion coefficient (with $\alpha =1$ defined to be normal diffusion, i.e.\ 
$\bD(t)$ being a constant).

We can generalize our results further by adding an interaction potential to the right-hand side 
of our 3D (or N-dimensional actually) temporal Fokker-Planck equation: 
\be
\frac{\pd \rho(\bx,t)}{\pd t}=\onehalf\bD(t)\cdot\nabla^2\rho(\bx,t)-\bV(t)\cdot\nabla\rho(\bx,t)-{\cal U}(\bx,t)\,\rho(\bx,t)\;.
\label{eq:fp3dgen}
\ee
This is an example of a Euclidean Schr\"{o}dinger equation of path integral fame (with an additional {\it drift} term) which  
can be matched to a large number of physical diffusive processes. 
Note that ${\cal U}(\bx,t)$ does not depend on spatial derivatives 
(as in a generalization that includes $\nabla^3$, $\nabla^4$, etc.\ or other velocity-type operators has not been included), 
therefore this first term proportional to the anisotropic diffusion coefficient, $\bD(t)$, is the only 
dissipative term of the equation. Also note well that this equation is linear and therefore does not 
suffer from the sensitivity to initial conditions problem, but yet as shown in the previous sections, intermittent phenomena 
are contained in solutions to this equation: As the temporal power-law dependence of $\bD(t)$ is adjusted, sub- and super-diffusive 
phenomena appear including perhaps surprisingly turbulent diffusion (see Figure~\ref{fig:anomdiff}) even though the defining equation of 
the analytic model of the current paper is linear [Eq.~(\ref{eq:fp3d})]. 

\begin{acknowledgments}
The author would like to thank David Krout, Bob Miyamoto, and 
Mike Shlesinger for discussions. 
This work was supported by the Office of Naval Research.
\end{acknowledgments}


\begin{thebibliography}{99}

\bibitem{Krout1} D.\ W.\ Krout, {\it Intelligent Ping Sequencing for Multiple Target 
Tracking in Distributed Sensor Fields}, Ph.D.\ diss., University of Washington, 2006.

\bibitem{Shlesinger} M.\ F.\ Shlesinger, {\it Search research}, Nature {\bf 443} (2006) 281--282.

\bibitem{rwbio} E.\ A.\ Codling, M.\ J.\ Plank, and S.\ Benhamou, 
{\it Review: Random walk models in biology}, 
J.\ R.\ Soc.\ Interface {\bf 5} (2008) 813--834.

\bibitem{supradiff} M.\ F.\ Shlesinger, 
{\it Supra-diffusion}, in {\it Processes with Long-Range Correlations: Theory and Applications}, 
eds.\ G.\ Rangarajan and M.\ Ding (Springer, Berlin, 2003), pp.\ 139--147.

\bibitem{anomdiff} R.\ Metzler and J.\ Klafter, 
{\it The random walk's guide to anomalous diffusion: a fractional dynamics approach},
Physics Reports {\bf 339} (2000) 1--77.

\end{thebibliography}
\end{document}